# An exponential random graph modeling approach to creating group-based representative whole-brain connectivity networks

Sean L. Simpson[a,*], Malaak N. Moussa[b], Paul J. Laurienti[c]

[a] *Department of Biostatistical Sciences, Wake Forest University School of Medicine Winston-Salem, NC, USA*
[b] *Neuroscience Program, Wake Forest University School of Medicine Winston-Salem, NC, USA*
[c] *Department of Radiology, Wake Forest University School of Medicine Winston-Salem, NC, USA*

*Corresponding author. Department of Biostatistical Sciences, Wake Forest University School of Medicine, Medical Center Boulevard, Winston-Salem, NC 27157, USA. Fax: +1 336 716 6427.
E-mail address: slsimpso@wakehealth.edu (S. Simpson).

ABSTRACT

Group-based brain connectivity networks have great appeal for researchers interested in gaining further insight into complex brain function and how it changes across different mental states and disease conditions. Accurately constructing these networks presents a daunting challenge given the difficulties associated with accounting for inter-subject topological variability. Viable approaches to this task must engender networks that capture the constitutive topological properties of the group of subjects' networks that it is aiming to represent. The conventional approach has been to use a *mean* or *median* correlation network (Achard et al., 2006; Song et al., 2009) to embody a group of networks. However, the degree to which their topological properties conform with those of the groups that they are purported to represent has yet to be explored. Here we investigate the performance of these *mean* and *median* correlation networks. We also propose an alternative approach based on an exponential random graph modeling framework and compare its performance to that of the aforementioned conventional approach. Simpson et al. (2011) illustrated the utility of exponential random graph models (ERGMs) for creating brain networks that capture the topological characteristics of a single subject's brain network. However, their advantageousness in the context of producing a brain network that "represents" a group of brain networks has yet to be examined. Here we show that our proposed ERGM approach outperforms the conventional *mean* and *median* correlation based approaches and provides an accurate and flexible method for constructing group-based representative brain networks.

## Introduction

Whole-brain connectivity analysis is a burgeoning area in neuroscience which is gaining prominence due to the need to understand how various regions of the brain interact with one another. The application of network and graph theory has facilitated these analyses and enabled examining the brain as an integrated system rather than a collection of individual components (Bullmore and Sporns, 2009). Despite the utility of network science in providing insight into the infrastructural properties of a given subject's brain, capturing and understanding these properties in a group of subjects presents challenges that impede research focusing on changes in complex brain function across different cognitive and disease states.

As noted in Rubinov and Sporns (2010), comparing brain networks across subjects and groups of subjects necessitates the development of accurate statistical comparison tools. Despite this need, the amount of work done in this area has not been commensurate with its level of importance (van Wijk et al., 2010). Developing a systematic approach to capture the network characteristics from a group of subjects' brain networks has great appeal and would help to fill the gap in the group comparison literature. Evaluating individual networks and combining the information limits studies to exploring features that can be distilled into simple quantitative metrics such as the commonly used *clustering coefficient (C)* and *path length (L)*. While obtaining average measures of metrics such as these in a group is valuable, it only gives a global view of the system and does not capture the complex organization within a population of networks. However, a group-based representative brain connectivity network provides a graph that typifies the complex structure of a set of brain networks. These representative brain networks could serve as null networks against which other networks and network models could be compared, as visualization tools (Song et al., 2009), as a way to topologically map and assess a collection of networks (Achard et al., 2006), as a way to represent an individual's network based on several experimental runs, as a means to conduct group-based modularity analyses (Meunier et al., 2009a,b; Valencia et al., 2009; Joyce et al., 2010), and as an instrument for identifying hub/node types in a modularity analysis (functional cartography) (Joyce et al., 2010). In addition to their utility in the previously mentioned contexts, the most important use for these representative networks is in the nascent area of brain dynamics which requires representative group-based networks for simulating and assessing information flow on a network. Simulating dynamics on large groups of brain networks in order to understand how their topology supports brain activity is infeasible computationally, necessitating an accurate summary network (Jirsa et al., 2010). A representative network provides a central tendency for a set of complex systems that allows determining the type of dynamic information transfer a population of brain networks can support in various cognitive and disease states. Consequently, deeper insight can be gained on brain-behavior and brain-disease relationships. This needed reduction of many systems to a single system cannot be performed by simple manipulations of the average network metrics in a group. Also, as novel network metrics are developed whose computational burden increases with the number of subjects being analyzed, these representative networks will prove valuable. However, creating these group-based "representative" networks is a daunting challenge given the difficulties associated with accounting for the inter-subject topological variability.

There has been a lot of previous work on multivariate group models for functional networks; however, their foci have been different from those here. Varoquaux et al. (2010) developed a modeling approach to improve the estimation of a subject's network based on group time series data. Cecchi et al. (2009) proposed methods to extract features with high discriminative power from subject-level time series data. Ramsey et al. (2010) discussed difficulties in inferring causal relations from fMRI time series. Rădulescu and Mujica-Parodi (2009) applied principal component network analysis to time series data from a limited number of ROIs in the brain. Our goal here is *not* to fit models to time series data, but to fit them to already constructed binary network data. That is, the approaches we examine are independent of how connections are determined from time series data. We start from correlation matrices here, but some of the aforementioned approaches could be used to construct subject-level binary networks from time series data that could then be embedded within methods that produce a group representative network based on the individual networks. The goal of these representative network techniques is to take a set of already constructed networks and produce a group network that typifies the topological structure of the individual graphs. Hence, these methods are independent of how the initial subject-level networks are constructed.

Thus far, researchers have taken one of three general approaches to generating a group-based representative functional network. The most common method has been to take the mean of the functional connectivity matrices of the subjects in a group and threshold this group mean matrix to get a *mean* network (Achard et al., 2006; Meunier et al., 2009a,b; Valencia et al., 2009). Although this approach is intuitive and computationally straight forward, as with use of the mean in any context, the resulting network may be unduly influenced by one or more outlying functional connectivity values. Additionally, this approach is *edge-based* since the averaging is done across the individual entries of the connectivity matrices; and thus, it ignores the topological properties of each subject's network. Another similar approach taken by researchers is to take the median of the functional connectivity matrices of the subjects and threshold this group median matrix to get a *median* network (Song et al., 2009). While this approach provides more robustness to outlying connectivity values, it is still *edge-based* and ignores the topological dependence among edges within each subject's network. A third, and more contrasting approach, was taken by Meunier et al. (2009b) and Joyce et al. (2010). They used a *best subject* network to represent the group by assessing between subject differences in network organization (using an information-based measure and the Jaccard index respectively) and identifying the most representative subject in the sample. The network of the most representative subject is not likely to best capture all topological metrics simultaneously, but provides a balance among all of the variables. It would be ideal to have a method that incorporates the data from all subjects directly, like the *mean* and *median edge-based* network approaches, while also having each metric maximally optimized within the representative network rather than balancing properties by selecting a typical subject. Toward this end, we propose an approach to creating group-based representative networks that utilizes exponential random graph models (ERGMs), also known as p* models (Frank and Strauss, 1986; Wasserman and Pattison, 1996; Pattison and Wasserman, 1999; Robins et al., 1999). These models enable achieving an efficient representation of complex network data structures by modeling a network's global structure as a function of local topological features.

Simpson et al. (2011) showed the utility of ERGMs for simulating a brain network that retains the constitutive properties of a single subject's original network. However, their usefulness in producing a brain network that "represents" the topological characteristics of a group of networks has yet to be explored. Also, despite the frequent use of *mean* and *median* group-based networks, the amount to which their topological properties coincide with those of the groups that they are purported to represent has yet to be investigated. Here we make use of resting-state fMRI data from ten normal subjects to 1) examine how well the mean and median approaches capture important topological properties of the group and 2) compare their performances to our ERGM approach.

**Materials and methods**

*Data and network construction*

Our analysis included fMRI data from 10 normal subjects (5 female, average age 27.7 years old [4.7 SD]). For each subject, 120 images were acquired during 5 minutes of resting using a gradient echo echoplanar imaging (EPI) protocol with TR/TE=2500/40 ms on a 1.5 T GE twin-speed LX scanner with a birdcage head coil (GE Medical Systems, Milwaukee, WI). The acquired images were motion corrected, spatially normalized to the MNI (Montreal Neurological Institute) space and re-sliced to 4×4×5 mm voxel size using an in-house processing script based on SPM99 package (Wellcome Trust Centre for Neuroimaging, London, UK). The resulting images were not smoothed in order to avoid artificially introducing local spatial correlation (van den Heuvel et al., 2008). These subjects were part of a larger study with further details provided in Peiffer et al. (2009).

The first step in performing the network construction was to calculate Pearson partial correlation coefficients between the time courses of all node pairs adjusted for motion and physiological noises (see Hayasaka and Laurienti, 2010 for further details). These node time courses were obtained by averaging the voxel time courses in the 90 distinct anatomical regions (90 ROIs-Regions of Interest) defined by the Automated Anatomical Labeling atlas (AAL; Tzourio-Mazoyer et al., 2002). Three types of networks were then generated based on the resulting 90×90 correlation matrices. Unweighted, undirected *subject-specific* networks were created by thresholding the correlation matrices for each subject to yield a set of adjacency matrices ($\mathbf{A}_{ij}$) with 1 indicating the presence and 0 indicating the absence of an edge between two nodes. A group-based *mean* network was constructed by averaging the correlation matrices of the 10 subjects (i.e., averaging element (i,j) across matrices) and then thresholding the resulting matrix to yield a mean adjacency matrix. Similarly, a group-based *median* network was produced by computing the median of element (i,j) across the 10 correlation matrices and then thresholding the resulting matrix. All networks were defined (thresholded) so that the relationship (denoted by S) between the number of nodes n and the average node degree $K$ is the same across networks. In particular, the networks were defined so that $S = \log(n)/\log(K) = 2.8$. This relationship is based on the path length of a random network with n nodes and average degree $K$, and can be rewritten as $n = K^S$ (see Hayasaka and Laurienti, 2010 for further details on this thresholding approach and the reasons for choosing such a value).

*Model definition*

Exponential random graph models (ERGMs) have the following form (Handcock, 2002):

$$P_{\boldsymbol{\theta}}(\mathbf{Y}=\mathbf{y}) = \kappa(\boldsymbol{\theta})^{-1} \exp\{\boldsymbol{\theta}^T \mathbf{g}(\mathbf{y})\} \tag{1}$$

Here $\mathbf{Y}$ is an n×n (n nodes) random symmetric adjacency matrix representing a brain network from a particular class of networks, with $\mathbf{Y}_{ij} = 1$ if an edge exists between nodes i and j and $\mathbf{Y}_{ij} = 0$ otherwise. We statistically model the probability mass function (pmf) $\left(P_{\boldsymbol{\theta}}(\mathbf{Y}=\mathbf{y})\right)$ of this class of networks as a function of the prespecified network features defined by the p-dimensional vector $\mathbf{g}(\mathbf{y})$, where $\mathbf{y}$ is the observed network. This vector of explanatory metrics can contain any graph statistic (e.g., number of edges) or node statistic (e.g., brain location of the node). The goal in defining $\mathbf{g}(\mathbf{y})$ is to identify local metrics that concisely summarize the global (whole-brain) structure. A subset of statistically compatible metrics for ERGMs is defined in Table 1 (Morris et al., 2008). By statistically compatible we mean that the metrics do not generally induce degeneracy issues discussed in Rinaldo et al. (2009). These issues concern the shape of the estimated pmf (e.g., a pmf in which only a few graphs have nonzero probability) and can lead to lack of model convergence and unreliable results. The parameter vector $\boldsymbol{\theta} \in \mathbb{R}^p$ (which must be estimated), associated with $\mathbf{g}(\mathbf{y})$, quantifies the relative significance of each network feature in explaining the structure of the network after accounting for the contribution of all other network features in the model. More specifically, $\theta$ indicates the change in the log odds of an edge existing for each unit increase in the corresponding explanatory metric. If the $\theta$ value corresponding to a given metric is large and positive, then that metric plays a considerable role in explaining the network architecture and is more prevalent than in the null model (random network with the probability of an edge existing ($p$) = 0.5). Conversely, if the value is large and negative, then that metric still plays a considerable role in explaining the network architecture but is less prevalent than in the null model. The normalizing constant $\kappa(\boldsymbol{\theta})$ ensures that the probabilities sum to one. This approach allows representing the global network structure (**y**) by locally specified explanatory metrics (**g(y)**), and is capable of capturing several important topological properties of brain networks simultaneously (Simpson et al., 2011). The fitted parameter values $(\boldsymbol{\theta})$ can then be utilized to understand particular emergent behaviors of the network (how local features give rise to the global structure).

Estimation of the model parameters $\boldsymbol{\theta}$ is normally done with either Markov chain Monte Carlo maximum likelihood estimation (MCMC MLE) or maximum pseudo-likelihood estimation (MPLE) (van Duijn et al., 2009 contains details). Model fits with MPLE are much simpler computationally than MCMC MLE fits and afford higher convergence rates with large networks. However, properties of the MPLE estimators are not well understood, and the estimates tend to be less accurate than those of MCMC MLE. Here we employ MCMC MLE to fit the model in equation 1 given that there were no convergence issues, though convergence could become an obstacle in more spatially resolved networks (e.g., voxel-based). Hunter et al. (2008b) provides further details about this estimation approach which can be implemented in the statnet package (Handcock et al., 2008) for the R statistical computing environment which we used here.

*Representative network creation*

In order to create a group-based representative network via ERGMs, it is necessary to get a group-based summary measure of the fitted parameter values $(\boldsymbol{\theta})$ for all subjects. The first step in this process involves identifying the most important explanatory metrics (**g(y)**) for each subject's network as done in Simpson et al. (2011). They implemented a graphical goodness of fit (GOF) approach (Hunter et al., 2008a) to select the "best" metrics from the set of potential metrics listed by category in Table 2 for each of the 10 networks (10 subjects). These categories were chosen based on properties of brain networks that are regarded as important in the literature (Bullmore and Sporns, 2009). These metrics were chosen because they are analogous to typical brain network metrics (e.g., clustering coefficient ($C$)) but have been developed to be statistically compatible with ERGMs. Figure 1 exemplifies the calculation of GWESP, GWNSP, and GWDSP, on a six-node example network as these metrics are not classically used in graph analysis. The distribution of the unweighted analogues of these metrics (ESP, NSP, and DSP) is given for simplicity. The weighted versions simply sum the values of the distribution giving less weight to those with more shared partners. For this example we note that the network has 1 set of connected nodes with 0 shared partners ($ESP_0$), 5 sets with 1 shared partner ($ESP_1$), 2 sets with 2 shared partners ($ESP_2$), and 0 sets with 3 or 4 shared partners ($ESP_3$ and $ESP_4$). Further details on the metrics are provided in Table 1 and Morris et al. (2008). The $\tau$ parameters associated with GWESP, GWDSP, GWNSP, and GWD were set to $\tau = 0.75$ for reasons mentioned in Simpson et al. (2011). The metrics in bold font (Edges, GWESP, and GWNSP) in Table 2 were those contained in at least half ($\geq 5$) of these 10 best models/best sets of explanatory metrics. Examining the uniformity of the selected explanatory metrics across subjects in this way is important due to metric interdependencies. In other words, for example, the fitted parameter value $(\theta_{edge})$ associated with the Edge metric in a given subject's model is statistically dependent upon/accounts for all other fitted parameter values in the model. Thus, an appropriate statistical comparison or summary of $\theta_{edge}$ across subjects requires that the fitted ERGMs for all subjects contain the same set of metrics ($\mathbf{g}(\mathbf{y})$). A detailed exposition of the aforementioned model selection approach is presented in Simpson et al. (2011) and partially reproduced in the Appendix.

Given the metrics in bold font in Table 2, the second step in creating our group-based representative networks was to refit the networks of all 10 subjects with the "best" group model

$$P_{\boldsymbol{\theta}}(\mathbf{Y}=\mathbf{y})=\kappa(\boldsymbol{\theta})^{-1}\exp\{\theta_1 Edges + \theta_2 GWESP + \theta_3 GWNSP\}. \tag{2}$$

An example set of graphical GOF plots for the best group model fitted to a subject's network are given in Figure 2. For good-fitting models, the plot of the observed network should closely match that of the simulated networks. We can see that in this figure the model does a good job of capturing the geodesic (global efficiency), degree, and triad census (motifs) distributions, but does not do as well at capturing the shared partner (local efficiency) distribution. However, the model likely fits well enough to still justify conclusions about local efficiency (as is shown later in the results). The precise implications of these GOF plots for conclusions drawn from the model have yet to be thoroughly examined in the literature. The estimated parameter values $(\theta_1, \theta_2, \theta_3)$ for the best group model fitted to each subject are displayed in Table 3 along with the mean and median of those values across subjects. The third step in creating our group-based representative networks was to employ these mean $(\overline{\theta}_1, \overline{\theta}_2, \overline{\theta}_3)$ and median $(\widetilde{\theta}_1, \widetilde{\theta}_2, \widetilde{\theta}_3)$ values to simulate random realizations of networks from their corresponding probability mass functions (pmfs) below:

$$P_{\boldsymbol{\theta}}(\mathbf{Y}=\mathbf{y})=\kappa(\boldsymbol{\theta})^{-1}\exp\{\overline{\theta}_1 Edges + \overline{\theta}_2 GWESP + \overline{\theta}_3 GWNSP\} \tag{3}$$

$$P_{\boldsymbol{\theta}}(\mathbf{Y}=\mathbf{y})=\kappa(\boldsymbol{\theta})^{-1}\exp\{\widetilde{\theta}_1 Edges + \widetilde{\theta}_2 GWESP + \widetilde{\theta}_3 GWNSP\}. \tag{4}$$

Five *mean ERGM* and five *median ERGM* based networks were simulated based on equations 3 and 4 respectively. Additionally, five more networks were simulated from each probability mass function (in equations 3 and 4) where the degree distributions (distribution of the number of connections; see next section) were constrained to have the same distribution as subject 16 whom we deemed to be most representative of the group in terms of this metric. In other words, for these degree-constrained simulations, the probability mass functions in equations 3 and 4 were constrained so that only networks whose overall degree distribution was the same as subject 16's had a non-zero probability. This approach is equivalent to (though much more efficient than) simulating networks from the unconstrained pmfs in equations 3 and 4 as before, and then choosing only those simulated networks with the same degree distribution as subject 16. Accurately capturing the degree distribution of a set of networks proved difficult; thus, given its importance in conferring vital properties to brain networks, this constrained simulation approach was taken to examine whether we could fix the degree distribution while still being able to properly represent other topological properties of the group. In total, then, we had 20 potential group-based representative networks generated from these simulations.

*Network assessment*

Neurobiologically relevant network metrics were calculated for all subjects' networks, the *mean* and *median* correlation networks, and all ERGM derived mean and median simulated networks. Unless otherwise noted, these metrics were calculated and evaluated using in-house processing scripts. A more detailed review of the metrics used in these analyses can be found in the literature (Rubinov and Sporns, 2010).

A commonly used measure of network connectivity, *degree* ($K$), was determined for each network. The *degree* for each node ($k_i$) in a network was computed as the total number of functional links that were associated with a node $i$. The *characteristic path length* ($L$) is a measure of the functional integration within a network and was calculated using Dijkstra's algorithm (Dijkstra, 1959) in the MatlabBGL package (David Gleich; Stanford University, Standford, CA). It is equivalent to the average shortest path length in a network and was found by generating a matrix of the geodesic distances between all node pairs. However, in the case of isolated nodes and subgraphs this value is infinitely distant from the *largest network component* ($N_c$). Because of this, the harmonic mean of the geodesic distances was used to calculate $L$:

$$L = \frac{N(N-1)}{\sum_{i \neq j} \frac{1}{d_{ij}}} \qquad (5)$$

where $d_{ij}$ is equal to the harmonic mean of the geodesic distance between nodes $i$ and $j$ (Latora and Marchiori, 2001; Newman, 2002, 2003). *Global efficiency* ($E_{glob}$), which is the reciprocal of the *characteristic path length* (Latora and Marchiori, 2001), captures the level of distributive processing in a network. Unlike the *characteristic path length, global efficiency* is scaled and ranges in value from zero to one, where the latter represents maximal distributed processing.

*Clustering coefficient (C)* is defined as the fraction of triangles around a particular node in a network and is thus a measure of local network segregation (Watts and Strogatz, 1998). Akin to the *clustering coefficient* is *local efficiency* ($E_{loc}$). For each node in a network, this value represents the average local sub-graph efficiencies of its neighboring nodes (Latora and Marchiori, 2001). Like $E_{glob}$, it is a scaled value that ranges from zero to one. Nodes within a network with values closer to one are those with connections that are predominately local.

*Assortativity* ($R_{jk}$) captures the likelihood a node is connected with other like nodes in a network (Newman, 2002, 2003). In this study *assortativity* was based on similarity of node *degree*. Networks with values closer to 1 were defined as assortative and exhibited high-high and low-low degree connections. Values that were closer to -1 were indicative of disassortative networks with high-low degree connections.

*Statistical assessments of group-based networks*

The following analyses were done to compare original subject data to the *mean* and *median* correlation networks as well as the degree-constrained and unconstrained *mean ERGM* and *median ERGM* simulation-based networks. First, values from each of the 90-nodes of a subject's network were used to calculate whole-network means for each metric discussed in the previous subsection. The mean and median of these whole-network metric values were then computed across all subjects and compared with the corresponding metric values for the *mean* and *median* correlation networks and degree-constrained and unconstrained *mean ERGM* and *median ERGM* simulation-based networks. We also calculated the Euclidean distance between metric values of the networks in order to assess how well the representative networks captured all of the group mean and median topological values simultaneously. That is, we evaluated the square root of the squared distances between the group mean and median metric values and those of the *mean* and *median* correlation networks and degree-constrained and unconstrained *mean ERGM* and *median ERGM* simulation-based networks. Second, with the exception of *degree* and *assortativity*, the nodal cumulative distributions of each metric were generated for each network. These were made in order to evaluate which of the representative networks best captured the metric distributions of the subjects' data. Finally, to assess how well each representative network captured the nodal connectivity (*degree*) of the original subjects' networks, degree distributions were made. These distributions show the likelihood with which a particular node in a network displays a certain degree (Lima-Mendez and van Helden, 2009). Additionally, to visually assess whether or not representative networks could capture the overall topology of the original subjects' networks, Harel-Koren Fast Multiscale graphs (Harel and Koren, 2001) were plotted using NodeXL.

**Results**

We implemented the network assessment procedure delineated in the previous section for the 20 *mean ERGM* and *median ERGM* simulation-based representative networks and compared the results to those of the frequently used *mean* and *median* networks based on the subjects' correlation matrices. Table 4 shows the results of this assessment and comparison for the 10 unconstrained and 10 degree-constrained simulated networks. As evidenced by the results in this table, the 10 unconstrained *mean ERGM* and *median ERGM* simulation-based representative networks more accurately capture the group mean and median values for path length ($L$), size of the giant component ($N_c$), mean degree ($K$), and global efficiency ($E_{glob}$) than the corresponding *mean* and *median* correlation networks. There are minimal differences in accuracy for clustering coefficient ($C$) and local efficiency ($E_{loc}$), with the *mean* and *median* correlation networks only being decisively more accurate for assortativity ($R_{jk}$). When assessing the accuracy in capturing all of the group mean and median topological values simultaneously via the distance metric, all five of the unconstrained *mean ERGM* simulation-based networks outperform the *mean* correlation network, while four of the five *median ERGM* simulation-based networks outperform the *median* correlation network. However, since our ultimate goal is to select *a* network that is most representative of the group, only one simulation-based network need outperform the *mean* and *median* correlation networks for our approach to be useful. Thus, the anomalous underperforming *median ERGM* simulation-based network is of little relevance in our context. An examination of the degree distributions for the subjects' networks, unconstrained *mean ERGM* and *median ERGM* simulation-based networks, and *mean* and *median* correlation networks in Figure 3a illustrates the fact that the *mean* and *median* correlation networks tend to overestimate this distribution while the unconstrained *mean ERGM* and *median ERGM* simulation-based networks generally underestimate it.

Table 4 also displays the results for the 10 degree-constrained *mean ERGM* and *median ERGM* simulation-based networks. As was the case with the unconstrained networks, these degree-constrained networks also more accurately capture the group mean and median values for path length (L) (though 4 of the 10 simulations do a fairly poor job), size of the giant component ($N_c$), mean degree ($K$) (by construction), and global efficiency ($E_{glob}$) than the corresponding *mean* and *median* correlation networks. Moreover, they are also more faithful to the group mean and median values for local efficiency ($E_{loc}$). There are minimal differences in accuracy for clustering coefficient ($C$), with the *mean* and *median* correlation networks again only being decisively more accurate for assortativity ($R_{jk}$); nevertheless, the *mean ERGM* and *median ERGM* simulation-based networks qualitatively capture the assortative behavior of the brain networks, but just overestimate it quantitatively. As evidenced by the distance metric, the 10 degree-constrained *mean ERGM* and *median ERGM* simulation-based representative networks uniformly outperform the corresponding *mean* and *median* correlation networks. Additionally, we calculated the Euclidean distance over all metrics except the mean degree ($K$) in order to highlight the contribution of constraining the degree distribution (Table 4). The degree-constrained simulation-based networks maintain their uniform outperformance of the *mean* and *median* correlation networks, with the most accurate ones, degree-constrained *mean ERGM* network #5 and *median ERGM* network #4, still besting all of the unconstrained networks. Figure 3b depicts the degree distributions of the degree-constrained *mean ERGM* and *median ERGM* simulation-based networks along with those of the *mean* and *median* correlation networks and subjects' networks. By construction, the *mean ERGM* and *median ERGM* simulation-based networks have degree distributions which are all the same and that well represent (fall near the middle of) those of the group. Given this, and based on the results of Table 4, degree-constrained *mean ERGM* network #5 and *median ERGM* network #4 serve as the two best candidates for the group-based representative network.

In addition to capturing mean nodal properties as demonstrated in Table 4, it was also of interest to examine how well the best ERGM simulation-based representative networks (*mean ERGM* network #5 and *median ERGM* network #4) typified the distribution of these nodal properties. Figure 4 displays the nodal distributions of path length ($L$), clustering coefficient ($C$), global efficiency ($E_{glob}$), and local efficiency ($E_{loc}$) for the 10 subjects' networks, *mean* and *median* correlation networks, and the most representative ERGM networks. The ERGM networks preserve the distribution of the nodal properties extremely well, clearly outperforming the *mean* and *median* correlation networks. The Harel-Koren Fast Multiscale graphs (Harel and Koren, 2001) in Figure 5 help to visualize the networks in two dimensions and provide further visual evidence of the topological discrepancies between the most representative ERGM networks and the *mean* and *median* correlation networks. Most notably, the ERGM based networks have denser, more cohesive clusters than the *mean* and *median* correlation networks. This architectural difference has major implications for examining the dynamics of brain networks. These tighter clusters also seem characteristic of the individual subjects' Harel-Koren Fast Multiscale graphs which are shown in Supplementary Figure 1. Examining graphs such as these is important in brain network analysis as they aid in discerning whether there are qualitative differences in networks that are not being captured by the quantitative measures being used. In other words, if two networks have equivalent metric values ($C$, $L$, etc.) but their topology appears different in the Harel-Koren Fast Multiscale graphs, then it is possible that an important property is not being quantitatively measured.

In addition to the original 20 *mean ERGM* and *median ERGM* simulation-based representative networks, we also simulated another 50 degree-constrained *mean ERGM* networks in order to assess how much better the "best" representative network would be when drawn from a larger pool. The mean Euclidean distance of these 50 simulations was 4.53 with the best network having a distance of 0.31 (compared with 0.45 for the best network from the original set (Table 4)). As expected, selecting a network from a larger set of simulations leads to a better representative network. However, this network is only slightly better than the one from the original set. Thus, if an accurate representative network is found in a smaller set of simulations, there are diminishing returns with higher numbers of simulations. The optimal number of simulations to conduct will vary by context and assessment metrics.

**Discussion and Conclusion**

The construction of representative group-based networks is of paramount importance for brain network scientists. The need for these representative networks in current areas of research is well documented (Song et al., 2009; Achard et al., 2006; Meunier et al., 2009a,b; Valencia et al., 2009; Joyce et al., 2010; Jirsa et al., 2010), and their potential utility for future areas is promising. When examining complex systems it is frequently important to examine features that cannot be distilled into a single quantitative metric. For example, if one wants to determine the modular or community structure that represents the population this cannot readily be done by combining the communities of the individual subjects. Each subject's communities are complex and there is no methodology or algorithm available to produce a community representation that is typical of the group. Similarly, if one would like to evaluate how information flows on a network with certain characteristics, it is not feasible to model information flow on each individual network and then combine these flow patterns. Our approach produces a connectivity graph that provides a central tendency for a set of complex systems that allows determining the type of dynamic information flow a population of brain networks can support in various cognitive and disease states (Jirsa et al., 2010). Another possible use of a representative network would be to evaluate how assault and failure of nodes or edges alters the emergent properties of the system. Performing assaults on the networks from individual subjects can be done, but the assessment of the group outcomes must be limited to relatively simple metrics such as clustering and path length. It is unlikely that such metrics capture the vital changes that occur within the complex organization of the system. Planned work aimed at further refining our approach to explicitly incorporate anatomical information will also allow comprehensively assessing which brain areas are consistently connected across large populations of subjects.

This list of potential uses for a representative network is not exhaustive but points out several methods currently used to examine networks that would benefit from such a representative network. As the field of network science advances it is anticipated that new methodologies will frequently be designed to assess single realizations of a complex system. In fact, the vast majority of the methodological advances being produced are coming from fields like statistical physics. These investigators most often use networks that do not have multiple representations (like the internet or a particular social system). We must develop methods to capitalize on the advances that are occurring in network science if we intend to use them to study normal and abnormal brain function.

The work herein illustrates the utility of the ERGM framework for producing group-based representative networks that capture both important average topological properties and nodal distributions of those properties in a group of networks better than the commonly used *mean* and *median* correlation networks. Of the 20 networks simulated based on our approach, 19 outperformed the *mean* and *median* correlation networks as assessed by our distance metric, with the best two being an order of magnitude more accurate; though, again, only *one* network (the one with the smallest distance) would ultimately be chosen. The representative network produced within this framework has the potential to be further improved by simply increasing the number of simulations and thereby increasing the probability of producing a network with an even smaller distance value. Moreover, this approach can be further refined by developing and assessing additional explanatory metrics that may lead to a more descriptive group model (equation 2). For example, as mentioned previously, incorporating anatomical information will prove useful given that the approach accounts only for topological properties of the graphs and may not enable coming to any conclusions about specific sets of edges or nodes since the anatomical locations in the ERGM simulated networks are lost without Nodematch in the model.

The general failure of the *mean* and *median* correlation approach to create appropriate group-based networks is likely due to the fact that each edge is treated as independent from all others, thus ignoring the dependence structure of the networks. Additionally, the clear inability of these approaches to capture the path length of the individual networks may be the result of long distance connections having more variability across subjects than local connections. Thus, these long distance connections may simply average out and fall below the threshold value. The fact that the *mean* and *median* correlation networks have a larger average degree than the individual networks (Table 4) may also contribute to their shorter path length (and higher clustering). This larger average degree may be the result of local connections having less variability across subjects than other connections. Thus, these local connections may average above the threshold even though not all of the subjects have a given connection. A plausible alternative explanation for the disparate metric values is that the size of the giant component ($N_c$) may be driving the differences in the other metric values (Alexander-Bloch et al., 2010; Dorogovtsev et al., 2008). Whatever the mechanism may be that is driving the failure of the *mean* and *median* correlation approach to capture these network metrics, metric interdependencies make it likely that just one or two of them are driving the results.

The other approach used to construct group-based representative networks (mentioned in the Introduction) not contrasted in detail with our ERGM approach here selects the network of the "best subject" to represent the group (Meunier et al., 2009b; Joyce et al., 2010). While this method would produce equally good results for the data assessed here, it suffers from the conceptual drawback that it does not incorporate data from all other subjects. Thus, it would preclude comprehensively assessing which brain areas are consistently connected across large populations of subjects. Additionally, the network of the selected representative subject may accurately capture the mean and distribution of some metrics but be quite atypical for others in groups with large inter-subject variability. Although ERGM based networks could also suffer from this problem, it is less likely given that the network represents a central tendency based on the data from all subjects. That is, there may not be a single subject's network that is representative across all network measures, whereas ERGM based networks are more likely to be (and can be tailored to be) so.

There are several limitations of the ERGM approach that may serve as the focus of future work. The computational intensiveness of fitting ERGMs may preclude their use in the construction of group-based representative networks with more spatially resolved networks than those based on the AAL atlas due to convergence issues. It is important to note that our approach is ad hoc, and may not work in all cases. Also, the amount of programming work increases linearly with the number of subjects in the group since ERGMs must be fitted and assessed for each subject individually. However, in contexts where the method appears feasible, like the one presented here, our approach is clearly more appealing than the current methods. For situations in which the ERGM approach appears infeasible, the development of modified *mean/median* correlation approaches that incorporate network dependencies will prove useful.

Currently, our ERGM based approach provides an excellent method for creating group-based representative whole-brain connectivity networks that surpasses the methods in current use. Our approach affords a way to gain insight into the "average" complex internal connectivity structure of brains from a group and how this structure is altered in groups with various diseases and disorders. The generation of a representative network for a population will enable investigators to evaluate complex properties that are typical of the study population. These properties can give greater insight into the organization and function of the network than that achieved by relying on global metrics such as clustering and path length. To fully understand the brain it will be vital that we study the emergent properties that arise given the network topology. It is also important to be able to evaluate changes in emergent properties when the topology is specifically altered. The generation of population-based representative networks will enable comparing complex processes between study populations and evaluating how these processes change for various brain disorders.

**Appendix: Graphical GOF model selection procedure** (partially reproduced from Simpson et al., 2011)

Steps 1 and 2 aim to select the best Connectedness and Local Efficiency metrics respectively from the two options in Table 2. We opted to select one of each to avoid potential collinearity issues that may arise given that metrics within the same category attempt to capture the same properties. We started with selection of the Connectedness metric first as inclusion of both Edges and Two-Path in the model generally led to non-convergence. The location metric Nodematch was assessed last as we wanted to examine whether we could capture nodal information after accounting for topological structure. This set of steps merely provides a needed ad hoc procedure for metric selection in this context and is only one of many possible approaches. Fitting all possible models is generally infeasible (especially as more explanatory metrics are considered); thus, parsimonious approaches like the one outlined here are needed.

Step 1 – Selection of Connectedness metric.
Fit $P_{\boldsymbol{\theta}}(\mathbf{Y}=\mathbf{y}) = \kappa(\boldsymbol{\theta})^{-1} \exp\{\theta_1 Edges + \theta_2 GWESP + \theta_3 GWDSP + \theta_4 GWNSP + \theta_5 GWD\}$ and
$P_{\boldsymbol{\theta}}(\mathbf{Y}=\mathbf{y}) = \kappa(\boldsymbol{\theta})^{-1} \exp\{\theta_1 \textit{Two-Path} + \theta_2 GWESP + \theta_3 GWDSP + \theta_4 GWNSP + \theta_5 GWD\}$. Retain Connectedness metric (*Edges* or *Two-Path*) from the model with the better graphical GOF plots (Hunter et al., 2008a), which we will denote as *C*.

Step 2 – Selection of Local Efficiency metric.
Fit $P_{\boldsymbol{\theta}}(\mathbf{Y}=\mathbf{y}) = \kappa(\boldsymbol{\theta})^{-1} \exp\{\theta_1 C + \theta_2 GWESP + \theta_3 GWNSP + \theta_4 GWD\}$ and
$P_{\boldsymbol{\theta}}(\mathbf{Y}=\mathbf{y}) = \kappa(\boldsymbol{\theta})^{-1} \exp\{\theta_1 C + \theta_2 GWDSP + \theta_3 GWNSP + \theta_4 GWD\}$. Retain Local Efficiency metric (*GWESP* or *GWDSP*) from the model with the better graphical GOF plots, which we will denote as *LE*.

Step 3 – Graphical GOF selection.
Fit $P_{\boldsymbol{\theta}}(\mathbf{Y}=\mathbf{y}) = \kappa(\boldsymbol{\theta})^{-1} \exp\{\theta_1 C + \theta_2 LE + \theta_3 GWNSP + \theta_4 GWD\}$ and all 4 possible models containing three explanatory metrics (see Simpson et al., 2011). Compare the 5 models and

select the one with the best GOF plots. This model will be denoted as
$P_{\boldsymbol{\theta}}(\mathbf{Y}=\mathbf{y})=\kappa(\boldsymbol{\theta})^{-1}\exp\{\boldsymbol{\theta}_r^T\mathbf{g}_r(\mathbf{y})\}$.

Step 4 – Selection of final model.

Fit $P_{\boldsymbol{\theta}}(\mathbf{Y}=\mathbf{y})=\kappa(\boldsymbol{\theta})^{-1}\exp\{\boldsymbol{\theta}_r^T\mathbf{g}_r(\mathbf{y})+\theta_{r+1}Nodematch\}$ to determine whether adding information about nodal location in the brain in this manner leads to a "better" model. The final model is the one with the better GOF plots between $P_{\boldsymbol{\theta}}(\mathbf{Y}=\mathbf{y})=\kappa(\boldsymbol{\theta})^{-1}\exp\{\boldsymbol{\theta}_r^T\mathbf{g}_r(\mathbf{y})+\theta_{r+1}Nodematch\}$ and $P_{\boldsymbol{\theta}}(\mathbf{Y}=\mathbf{y})=\kappa(\boldsymbol{\theta})^{-1}\exp\{\boldsymbol{\theta}_r^T\mathbf{g}_r(\mathbf{y})\}$.

## Acknowledgements

This work was supported by the Translational Science Institute of Wake Forest University (Translational Scholar Award) and the National Institute of Neurological Disorders and Stroke (NS070917).

**Table 1**
Subset of ERGM explanatory metrics

| Metric | Description |
| --- | --- |
| Edges | Number of edges in the network |
| Two-Path | Number of paths of length 2 in the network |
| k-Cycle | Number of k-cycles in network |
| k-Degree | Number of nodes with degree k |
| Geometrically weighted degree (GWD) | Weighted sum of the counts of each degree ($i$) weighted by the geometric sequence $(1-\exp\{-\tau\})^i$, where $\tau$ is a decay parameter |
| Geometrically weighted edge-wise shared partner (GWESP) | Weighted sum of the number of connected nodes having exactly $i$ shared partners weighted by the geometric sequence $(1-\exp\{-\tau\})^i$, where $\tau$ is a decay parameter |
| Geometrically weighted non-edge-wise shared partner (GWNSP) | Weighted sum of the number of non-connected nodes having exactly $i$ shared partners weighted by the geometric sequence $(1-\exp\{-\tau\})^i$, where $\tau$ is a decay parameter |
| Geometrically weighted dyad-wise shared partner (GWDSP) | Weighted sum of the number of dyads[a] having exactly $i$ shared partners weighted by the geometric sequence $(1-\exp\{-\tau\})^i$, where $\tau$ is a decay parameter |
| Nodematch | Number of edges (i,j) for which nodal attribute i equals nodal attribute j (e.g., brain location of node i = brain location of node j) |

[a]node pair with or without edge

**Table 2**
Explanatory network metrics by category

| Category | Metric(s)[b] |
|---|---|
| 1) Connectedness | **Edges**, Two-Path |
| 2) Local Clustering/Efficiency | **GWESP**, GWDSP |
| 3) Global Efficiency | **GWNSP**[a] |
| 4) Degree Distribution | GWD |
| 5) Location (in the brain) | Nodematch |

[a]Not inherently global, but helps produce models that accurately capture the global efficiency of our networks

[b]Metrics in bold font were those contained in at least half of the "best" subject network models

**Table 3**
Group model parameter estimates

| Subject | $\theta_1$ (Edges) | $\theta_2$ (GWESP) | $\theta_3$ (GWNSP) |
|---|---|---|---|
| 002 | -2.562 | 0.966 | -0.293 |
| 003 | -1.842 | 0.745 | -0.329 |
| 005 | -2.789 | 1.016 | -0.279 |
| 008 | -3.367 | 1.311 | -0.266 |
| 009 | -3.287 | 1.092 | -0.180 |
| 010 | -3.666 | 1.335 | -0.204 |
| 012 | -2.549 | 0.973 | -0.266 |
| 013 | -2.791 | 1.044 | -0.301 |
| 016 | -2.525 | 0.951 | -0.308 |
| 021 | -2.316 | 0.704 | -0.365 |
| **Mean** | -2.769 | 1.014 | -0.279 |
| **Median** | -2.676 | 0.994 | -0.286 |

**Table 4**
Network assessment results for the subjects' networks, edge-based *mean* and *median* correlation networks, and *mean ERGM* and *median ERGM* simulation-based networks.

| | | $N_c$ | L | K | C | $E_{loc}$ | $E_{glob}$ | R | | |
|---|---|---|---|---|---|---|---|---|---|---|
| **Subject** | | | | | | | | | | |
| Mean | | 83.7 | 4.14 | 5.05 | 0.40 | 0.49 | 0.28 | 0.29 | | |
| Median | | 84.5 | 4.18 | 5.04 | 0.41 | 0.49 | 0.28 | 0.30 | | |
| **Edge-Based Network** | | | | | | | | | *Distance with K* | *Distance without K* |
| Mean | | 89 | 3.54 | 6.42 | 0.44 | 0.55 | 0.35 | 0.33 | **5.51** | **5.33** |
| Median | | 88 | 3.01 | 7.60 | 0.45 | 0.59 | 0.39 | 0.22 | **4.49** | **3.69** |
| **EGRM Simulation** | | | | | | | | | | |
| | | | | | *Unconstrained* | | | | | |
| Mean | 1 | 85 | 4.48 | 4.09 | 0.37 | 0.45 | 0.26 | 0.11 | **1.66** | **1.36** |
| | 2 | 85 | 3.63 | 4.58 | 0.35 | 0.44 | 0.30 | 0.09 | **1.49** | **1.41** |
| | 3 | 83 | 4.12 | 3.76 | 0.39 | 0.45 | 0.26 | 0.12 | **1.48** | **0.72** |
| | 4 | 88 | 4.28 | 4.27 | 0.45 | 0.52 | 0.29 | 0.02 | **4.38** | **4.31** |
| | 5 | 83 | 4.59 | 3.84 | 0.40 | 0.47 | 0.24 | 0.20 | **1.47** | **0.84** |
| Median | 1 | 86 | 3.93 | 4.62 | 0.34 | 0.41 | 0.30 | 0.40 | **1.58** | **1.53** |
| | 2 | 87 | 4.48 | 4.31 | 0.42 | 0.51 | 0.27 | 0.20 | **2.63** | **2.52** |
| | 3 | 86 | 4.35 | 4.64 | 0.42 | 0.49 | 0.28 | 0.39 | **1.57** | **1.51** |
| | 4 | 86 | 3.83 | 4.71 | 0.36 | 0.46 | 0.30 | 0.33 | **1.58** | **1.54** |
| | 5 | 78 | 4.72 | 3.78 | 0.43 | 0.49 | 0.22 | 0.20 | **6.65** | **6.52** |
| | | | | | *Constrained* | | | | | |
| Mean | 1 | 83 | 4.43 | 5.04 | 0.45 | 0.56 | 0.26 | 0.47 | **0.79** | **0.79** |
| | 2 | 83 | 5.51 | 5.04 | 0.43 | 0.51 | 0.23 | 0.61 | **1.57** | **1.57** |
| | 3 | 81 | 6.35 | 5.04 | 0.46 | 0.56 | 0.21 | 0.42 | **3.50** | **3.50** |
| | 4 | 83 | 4.26 | 5.04 | 0.40 | 0.49 | 0.27 | 0.55 | **0.76** | **0.76** |
| | 5 | 84 | 4.30 | 5.04 | 0.34 | 0.45 | 0.27 | 0.57 | **0.45** | **0.45** |
| Median | 1 | 86 | 5.21 | 5.04 | 0.39 | 0.48 | 0.25 | 0.59 | **1.85** | **1.85** |
| | 2 | 86 | 4.32 | 5.04 | 0.34 | 0.43 | 0.28 | 0.62 | **1.54** | **1.54** |
| | 3 | 83 | 5.17 | 5.04 | 0.44 | 0.52 | 0.24 | 0.46 | **1.81** | **1.81** |
| | 4 | 84 | 4.16 | 5.04 | 0.37 | 0.47 | 0.28 | 0.49 | **0.54** | **0.54** |
| | 5 | 83 | 4.63 | 5.04 | 0.40 | 0.51 | 0.26 | 0.47 | **1.57** | **1.57** |

Note: "Distance" denotes the Euclidean distance between metric values of the potential representative networks and the corresponding mean or median metric values of the subjects' networks.
"Distance w/o K" denotes the same distance with mean degree (K) left out of the calculation.

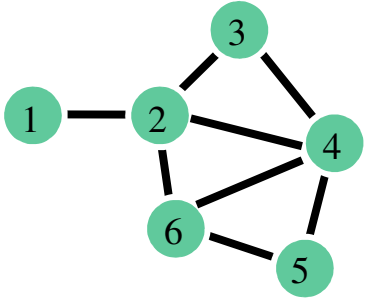

**Fig. 1.** Six-node example network (reproduced from Simpson et al., 2011). The edgewise, nonedgewise, and dyadwise shared partner distributions are $(ESP_0,\ldots, ESP_4) = (1, 5, 2, 0, 0)$, $(NSP_0,\ldots, NSP_4) = (1, 4, 2, 0, 0)$, and $(DSP_0,\ldots, DSP_4) = (2, 9, 4, 0, 0)$ respectively.

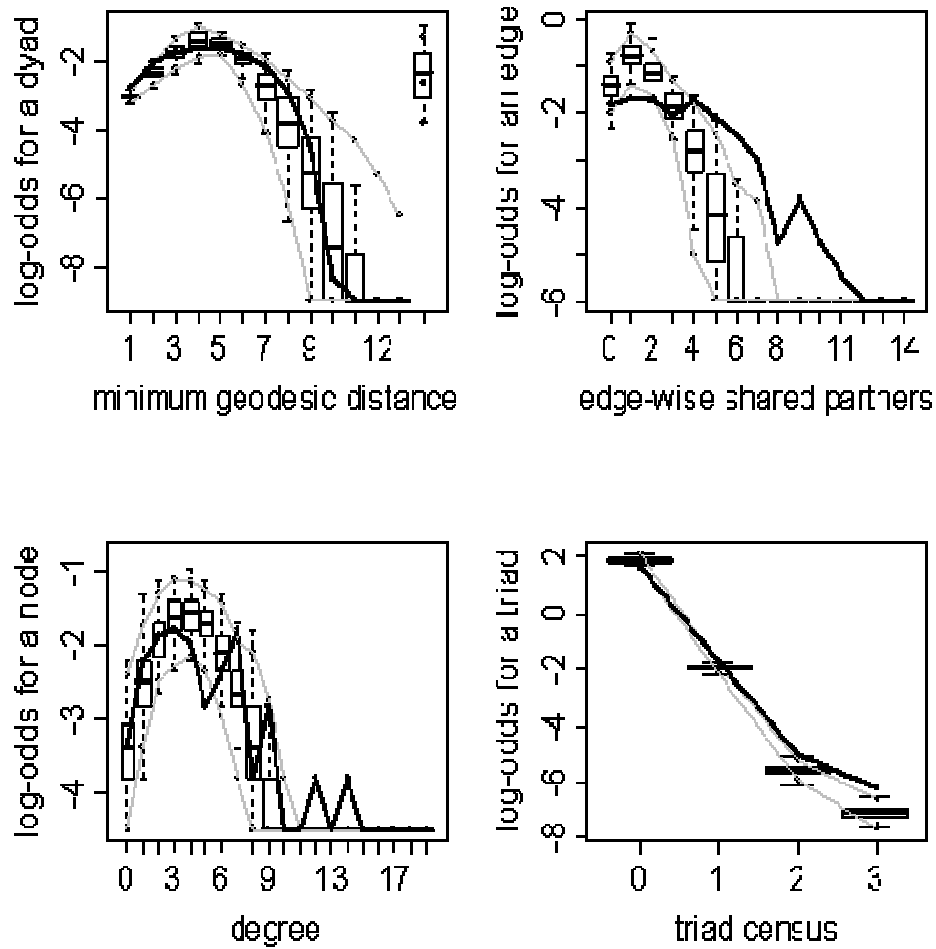

**Fig. 2.** Goodness-of-fit plots for the best group model fitted to subject 5. The vertical axis is the logit of relative frequency, the solid lines represent the statistics of the observed network, and the boxplots represent the distributions of the 100 simulated networks.

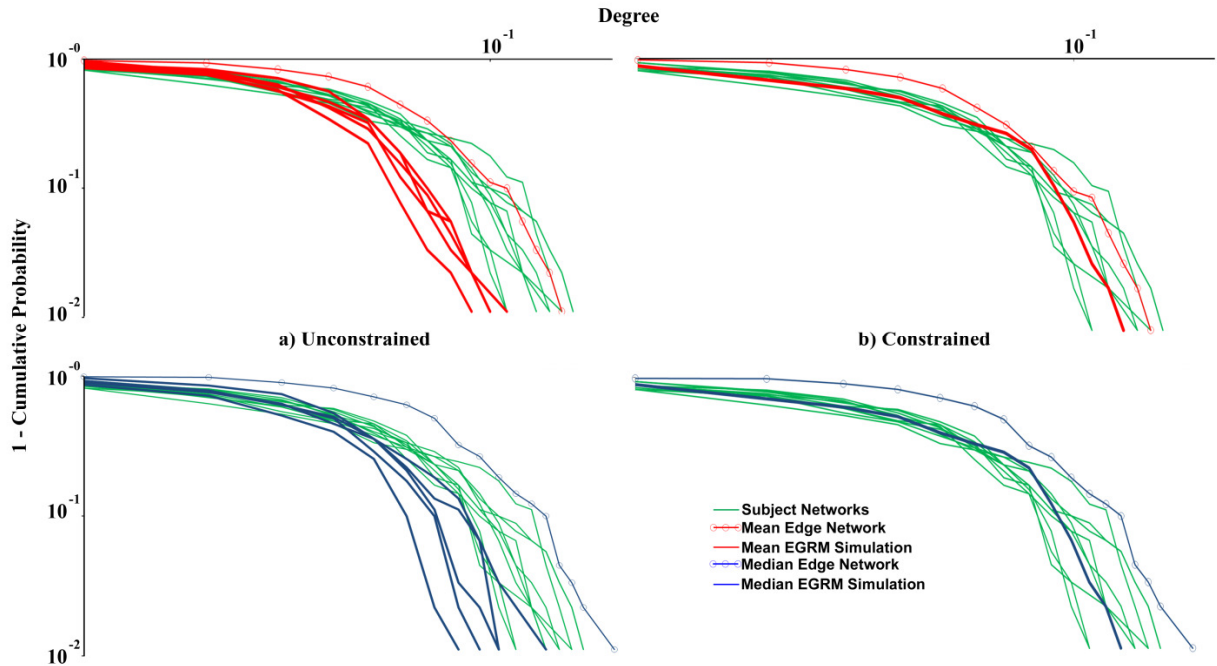

**Fig. 3.** Degree distribution plots for the 10 subjects' networks, the edge-based *mean* and *median* correlation networks, and **a)** the unconstrained *mean ERGM* and *median ERGM* simulation-based networks and **b)** the degree-constrained *mean ERGM* and *median ERGM* simulation-based networks. The upper panels show the *mean* networks and the lower panels the *median* networks.

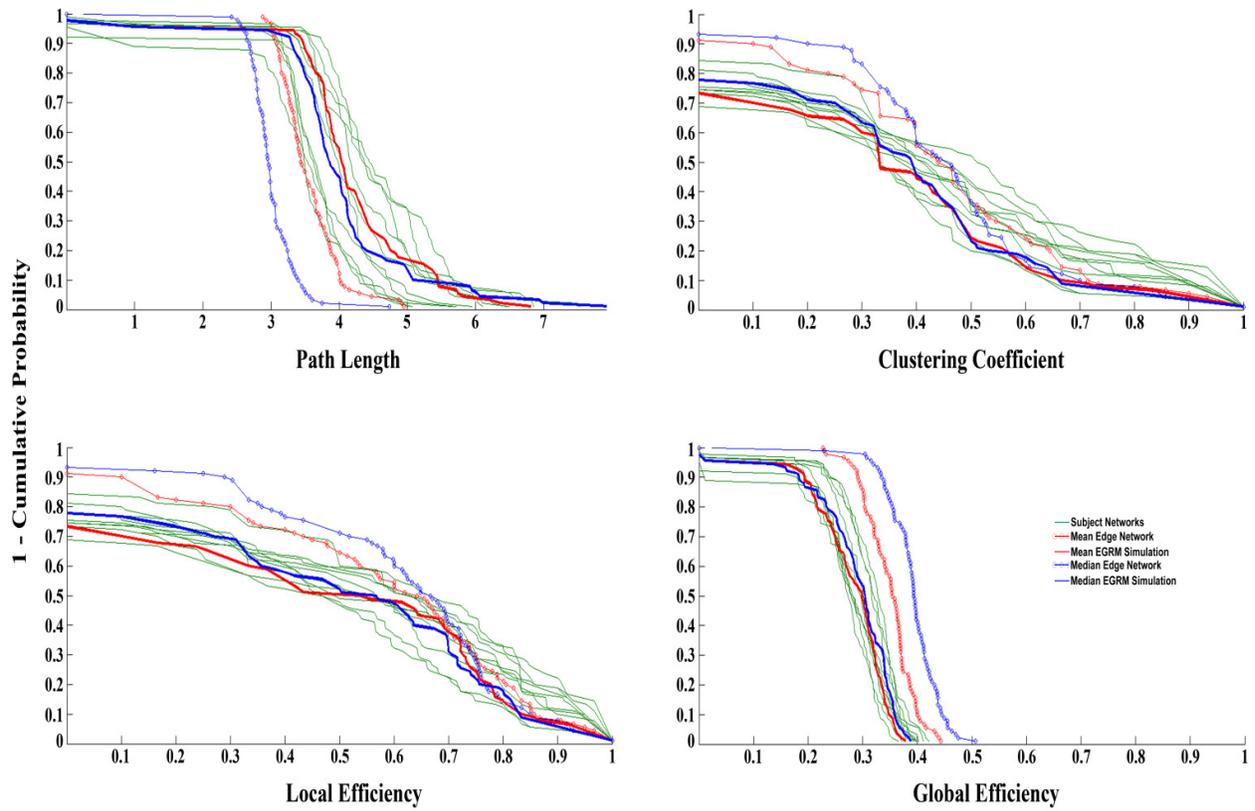

**Fig. 4.** Nodal distributions of path length ($L$), clustering coefficient ($C$), global efficiency ($E_{glob}$), and local efficiency ($E_{loc}$) for the 10 subjects' networks, *mean* and *median* correlation networks, and degree-constrained *mean ERGM* network #5 and *median ERGM* network #4.

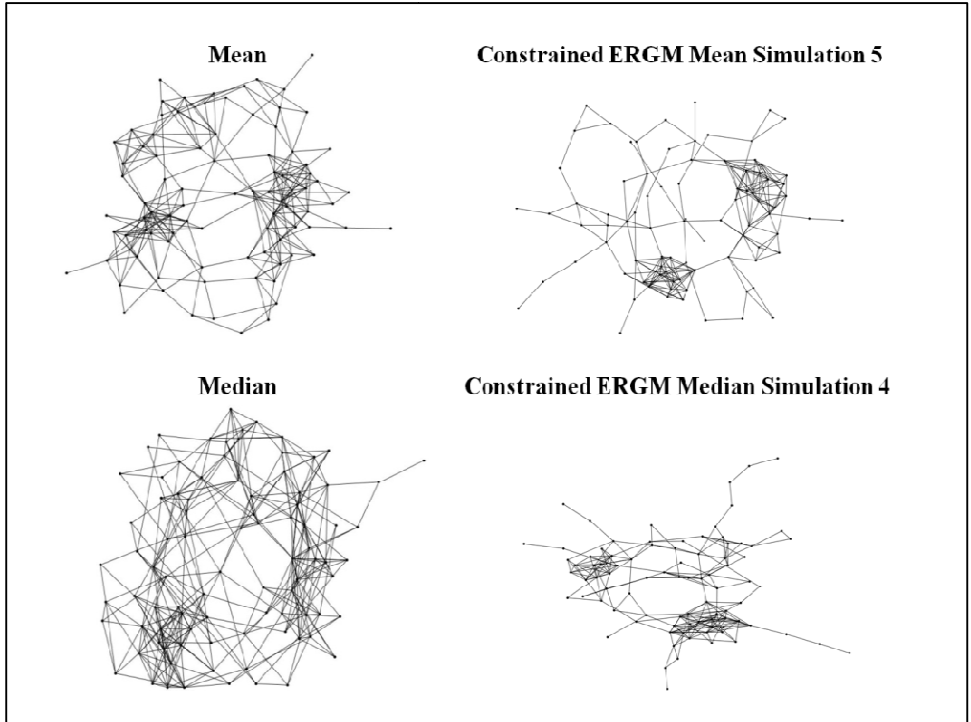

**Fig. 5.** Harel-Koren Fast Multiscale graphs for the *mean* and *median* correlation networks, degree-constrained *mean ERGM* network #5, and degree-constrained *median ERGM* network #4.